# Revealing photons' past via quantum twisted double-slit experiments

Zhi-Yuan Zhou[1,2], Zhi-Han Zhu[3,*], Shi-Long Liu[1,2], Yin-Hai Li[4], Shuai Shi[1,2], Dong-Sheng Ding[1,2], Li-Xiang Chen[5], Wei Gao[3], Guang-Can Guo[1,2], and Bao-Sen Shi[1,2,*]

[1]CAS Key Laboratory of Quantum Information, USTC, Hefei, Anhui 230026, China.
[2]Synergetic Innovation Center of Quantum Information & Quantum Physics, University of Science and Technology of China, Hefei, Anhui 230026, China.
[3]Da-Heng Collaborative Innovation Center for Science of Quantum Manipulation & Control, Harbin University of Science and Technology, Harbin 150080, China.
[4]Department of Optics and Optical Engineering, University of Science and Technology of China, Hefei, Anhui 230026, China.
[5]Department of Physics and Collaborative Innovation Center for Optoelectronic Semiconductors and Efficient Devices, Xiamen University, Xiamen 361005, China.
*e-mail: zhuzhihan@hrbust.edu.cn; drshi@ustc.edu.cn;

**Are quantum states real[1-4]? This most fundamental question in quantum mechanics has not yet been satisfactorily resolved, although its realistic interpretation seems to have been rejected by various delayed-choice experiments[5]. Here, to address this long-standing issue, we present a quantum twisted double-slit experiment. By exploiting the subluminal feature of twisted photons, the real nature of a photon during its time in flight is revealed for the first time. We found that photons' arrival times were inconsistent with the states obtained in measurements but agreed with the states during propagation. Our results demonstrate that wavefunctions describe the realistic existence and evolution of quantum entities rather than a pure mathematical abstraction providing a probability list of measurement outcomes. This finding clarifies the long-held misunderstanding of the role of wavefunctions and their collapse in the evolution of quantum entities.**

Over the development of quantum mechanics in the past century, a constant debate has continued as to whether wavefunctions derived in theory or measured in experiments describe the reality of quantum entities' existence and dynamic trajectory[1-7]. In the single-photon double-slit experiment dedicated to this problem, Copenhagen (standard) interpretation claims that the photons passing through the double slits have no definite nature until they are measured and that wavefunctions provide only 'a catalog of knowledge'. In contrast, the determinists (common-sense understanding) argue that the past of photons should be realistic and deterministic prior to the detection. To investigate when a photon's nature is determined and whether there are 'hidden variables', a number of delayed-choice experiments have been proposed and conducted over the past several decades[5]. The results show that the detecting device always determines which nature of a photon can be observed—as a wave, a particle or even a wave-particle superposition[8-20]. Based on the thinking paradigm that the past can be deduced from time reversal of the observed phenomenon, these results lead to the time paradox of a choice made in the present to alter a photon's past.

Therefore, the Copenhagen interpretation has to deny the reality of wavefunctions and follows a creed that *no elementary phenomenon is a phenomenon until it is a registered phenomenon*, making quantum mechanics an epistemic theory. Even more puzzling, the recently observed wave-particle superposition state has led researchers to question Bohr's principle of complementarity[19-20]. Moreover, a disconnected trace left by photons in a Mach–Zehnder interferometer, through which photons could not pass, was observed[21]. These surprising findings make the long-standing debate more intense, and the core of the problem is that if wavefunctions are not real, do the results measured at a given moment represent the photon's physical reality in the past?

Additionally, light travelling at a speed of $c$ in vacuum is another fundamental principle in modern physics that has also been well-known for a century. For this principle, however, the wave-particle duality will lead to a dual description of light speed, i.e., phase velocity $v_{ph}$ and group velocity $v_g$, corresponding to the propagation speed of the wavefront and the energy (not information), respectively. For a true plane-wave light field in vacuum, $v_{ph} = v_g = c$ due to the perfect plane wavefront without any transverse component of the wave vector. For a physically realizable light field, in contrast, the inevitable transverse structure of the wavefront leads to a reduction of the wave vector in the axial component that will slightly reduce the group velocity, which has recently fascinated the optical physics community[22-23]. Even more noteworthy, the group velocity dispersion that occurs in free space enables one to extract the propagation behavior of individual photons on their way to the detector. In this work, we present a quantum twisted double-slit experiment: photonic orbital angular momentum (OAM) degree of freedom is employed to 'mark' the double slits, causing photons that come from different slits to arrive at different times. This variation in arrival time provides an interface for investigating photons' propagation history. The results indicate that individual photons travel along both paths after passing the double slits, just as described by the wavefunction (definitely behaving as a wave). Therefore, the wavefunction describes a realistic manner for a photon as it travels to the detector rather than a pure mathematical abstraction.

Laguerre–Gaussian (LG) modes are natural eigenstates of OAM, which can form an infinite-dimensional Hilbert space[24-26]. As a type of physically realizable light field, the relationship between the wave vector $k_0$ and related components in cylindrical coordinates $\{r, \varphi, z\}$ can be expressed as $k_0^2 = k_r^2 + k_\varphi^2 + k_z^2$; therefore, the axial-group velocity can be expressed as $v_{g,z} = ck_z(k_0)^{-1}$. $v_{g,z}$ is always less than $c$, and it decreases with the azimuthal indices $\ell$. Furthermore, the twisted light generated by a spatial light modulator (SLM) is not a pure LG mode, but is a superposition of different LG modes in the form of a hypergeometric function (see theoretical frame in Methods and *Supplementary Materials*).[27]

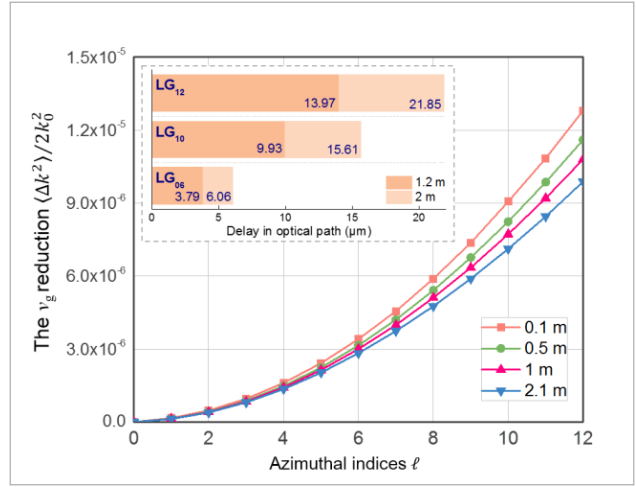

**Figure 1 | Calculated decrease in $v_{g,z}$ for twisted photons generated from a SLM.** Reduction in twisted photons' group velocity as a function of the azimuthal indices $\ell$ at different propagation distances after generation. The results are calculated from a realistic wavefunction of a light field in the form of a hypergeometric function generated from a SLM (for more details, see the theoretical calculation in *Supplementary Materials*). The upper left inset presents the accumulated delays of different LG modes after propagating 1.2 and 2 m.

Figure 1 shows the theoretical reduction of $v_{g,z}$ calculated from this hypergeometric twisted light in ideal conditions versus the propagation distance after generation. Obviously, the amount of reduction increases with $\ell$, which leads to an accumulated delay in the optical path obeying Eq. 3, as shown in the Methods section. Specifically, after propagating 1.2 m and 2 m, the accumulated delays $\tau_\ell(z)$ are approximately 3.8 μm and 6 μm for $\ell = 6$, 10 μm and 15.6 μm for $\ell = 10$, and 14 μm and 21.9 μm for $\ell = 12$, respectively. For a superposition state $|\varphi_{0\ell}(t)\rangle = \alpha|0(t)\rangle + \beta|\ell(t)\rangle$, where $\alpha$ and $\beta$ are mode weight coefficients (MWCs) obeying a relationship of $|\alpha|^2 + |\beta|^2 = 1$, the accumulated delay, according to the principle of linear superposition, will manifest as a linear superposition of the Gaussian and LG modes (paths) obeying Eq. 6, as shown in Methods, i.e., $|\beta|^2 \tau_\ell(z)$. Note that the accumulated delay depends only on the propagation history (propagating distance and spatial mode) before detection. Thus, the OAM and the $v_{g,z}$ reduction are employed to provide the which-slit selective interface and the propagation history information of individual photons, respectively.

In this work, the OAM is used to label the outer slit of the double slits, as depicted in the schematic diagram of the experimental setup shown in Fig. 2 (for details, see Methods and *Supplementary Materials*). The twisted double slits are mimicked by a spatial light modulator (SLM), which loads a phase mask of an OAM superposition state $|\varphi_{0\ell}\rangle = \alpha|0\rangle + \beta|\ell\rangle$. Here, $\alpha$ and $\beta$ (MWCs) depend on both the waist of the incident photons and the diameter of the Gaussian slit in the grayscale image. The transverse spatial wavefunction of the photons after passing this apparatus is $|\varphi_{0\ell}(t)\rangle$, as depicted in Fig. 2(b).

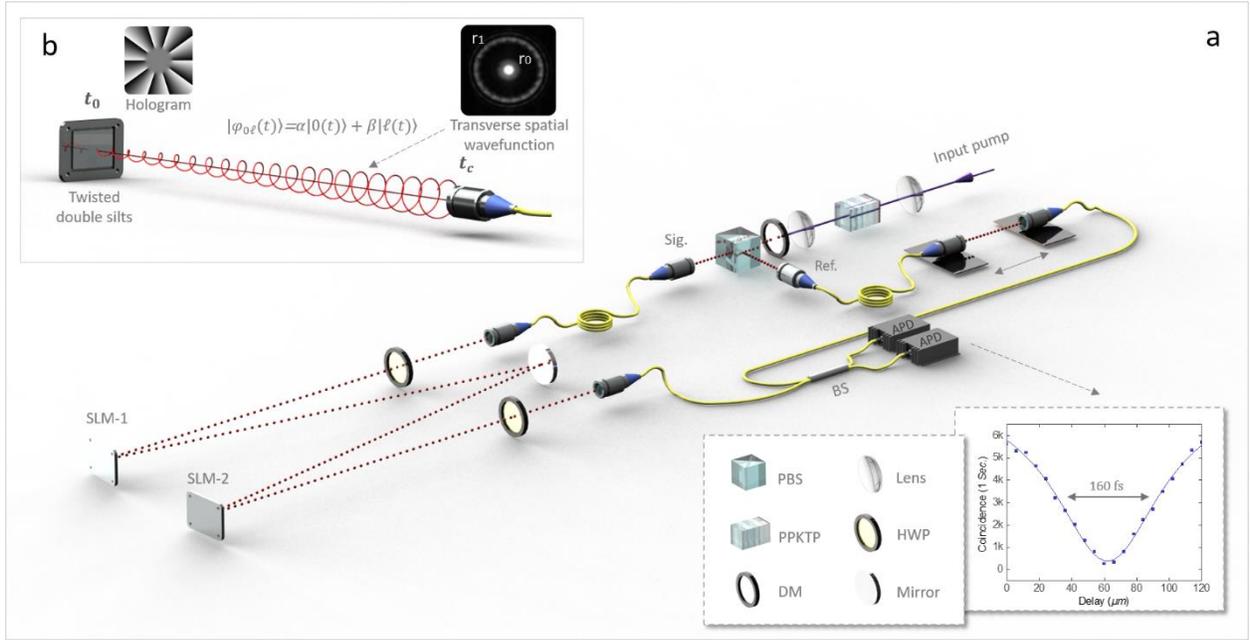

**Figure 2 | Schematic illustration of the quantum twisted double-slit experiment. a,** Experimental setup. A periodically poled potassium titanyl phosphate (PPKTP) crystal is pumped by violet light to create degenerate orthogonally polarized photon pairs at 795 nm. The photon pairs are then injected into two single-mode fibers by passing a dichroic mirror (DM) and a polarizing beam splitter (PBS). One photon is used as a herald photon travelling through the reference path, which contains a fiber link and an adjustable air gap. The other photon is directed into a free-space link consisting of two SLMs and is then transmitted back to the fiber and used as a signal, where two half-wave plates (HWPs) are adopted to match the maximum efficiency of the SLMs; the two SLMs provide different propagation distances. Finally, signal and reference arms are connected via a beam splitter (BS) and then directed into two single-photon avalanche photodiodes (APDs) to perform a coincidence measurement. **b,** The upper left inset shows a detailed schematic of photon propagation after passing the twisted double slit. After the transformation of the SLM with a superposition state $|\varphi_{0\ell}\rangle = \alpha|0\rangle + \beta|\ell\rangle$, the MWC (probability of photons in each mode) depends on both the radius of the Gaussian region in the hologram and the waist of the incident photons.

Consider the radius of the maximum intensity of LG modes $r_1 = \sqrt{|\ell|/2} \cdot w(z)$, where $w(z)$ is the beam radius upon propagation[28]; the radius of the LG modes increases with $\ell$ and the propagation distance. This result indicates that the transverse profile of photons passing the double slits is $\langle \varphi_{0\ell} | \varphi_{0\ell} \rangle = \alpha^2 \langle 0|0\rangle + \beta^2 \langle \ell|\ell\rangle + \alpha^*\beta \langle 0|\ell\rangle + \alpha\beta^* \langle \ell|0\rangle$, i.e., a Gaussian part $\alpha^2 \langle 0|0\rangle$ distributes around the optical axis $r_0$, an LG part $\beta^2 \langle \ell|\ell\rangle$ distributes at the outer ring $r_1$, and an interference part $\alpha^*\beta \langle 0|\ell\rangle + \alpha\beta^* \langle \ell|0\rangle$ lies between them. This setup is similar to Young's apparatus, with the difference that the photons passing the LG slit will slow down. If we place a screen after the double slits, i.e., choosing wave nature, the interference pattern $\langle \varphi_{0\ell} | \varphi_{0\ell} \rangle$ will emerge on the screen, and an 'intermediate delay' manifesting as a linear superposition of those coming from both paths (modes) obeying Eq. 6 can be observed. Once we selectively receive photons via Gaussian-mode projection (collapse measurement), i.e., a which-path setup for observing particle nature in the $|0\rangle$ path, the wavefunction $|\varphi_{0\ell}(t)\rangle$ will collapse into $|0(t)\rangle$ with a probability of $\alpha^2$ at the time of mode projection $t_c$, leading to further debate. Specifically, compared with a Gaussian mode (SLM working in mirror mode), there should be no delay in the photons' arrival time if the measured state ($|0\rangle$) represents the photons' physical reality in the past; the 'intermediate delay' will still emerge if the history of free-space propagation had been formed by the wavefunction $|\varphi_{0\ell}(t)\rangle$ before the collapse measurement at time $t_c$. Here, we use this 'paradox in arrival time' to verify which result represents the photons' real past.

In the experiment, as shown in Fig. 2(b), the 'screen' is always absent and we receive only photons that appear to have followed a Gaussian path via a single-mode fiber (SMF) directed toward the optical axis $r_0$ (mode distinguishability $D \approx 1$, for details, see *Supplementary Materials*). According to the Copenhagen viewpoint, the received photons would be chosen to behave as particles throughout the whole experimental apparatus, and the wavefunction $|\varphi_{0\ell}(t)\rangle$ can only tell us that the maximum probability of successfully receiving a photon is $\alpha^2$. The remaining question is whether the observed state truly represents the history of the photons. The answer is no; the Hong-Ou-Mandel (HOM) interference shown in Fig. 3 indicates that the arrival time of the signal photons received by Gaussian-mode projection still shows an 'intermediate delay', which is the theoretical prediction calculated from the corresponding wavefunctions $|\varphi_{0\ell}(t)\rangle$ (from $t_0$ to $t_c$). This result suggests that the 'trajectory' of the received photons does not follow the Gaussian path but both paths, or rather, it clearly behaves as a wave before being successfully received into the fiber. Specifically, individual photons in a superposition state $\sqrt{1/2}(|0\rangle+|12\rangle)$ are generated from SLM-1 and SLM-2, respectively, corresponding to a 2-m and 1.2-m free-space propagation distance. The results in Fig. 3(a) show measured delays of 4.93±0.61 μm for 1.2-m propagation and 7.51±0.46 μm for 2-m propagation; this result reveals that the wavefunction $|\varphi_{0\ell}(t)\rangle$ represents the past of the photons before they are received in the fiber (collapse into $|0(t)\rangle$). Moreover, the propagation distance is fixed at 2 m when SLM-2 operates in mirror mode, and we further verify the linear superposition principle under a realistic interpretation. Figure 3(b) shows the measured relative delay, which was 3.76±0.33 μm between $|0\rangle$ and $\sqrt{1/2}(|0\rangle+|6\rangle)$ and 4.12±0.56 μm between $\sqrt{1/2}(|0\rangle+|6\rangle)$ and $\sqrt{1/2}(|0\rangle+|12\rangle)$. Figure 3(c) shows the measured delay of the $|\varphi_{010}\rangle$ mode for different MWCs: a delay of 6.06±0.36 μm for $\sqrt{1/2}(|0\rangle+|10\rangle)$ and a delay of 3.29±0.51 μm for $\sqrt{3/4}|0\rangle+\sqrt{1/4}|10\rangle$. The delays of the received photons manifest a linear supposition of different modes and are consistent with the MWCs in corresponding wavefunctions, as indicated by the theoretical prediction of Eq. 6 in the Methods section. These results confirm that the transverse spatial wavefunctions shown in Fig. 2(b) and 3 describe realistic histories of the photons' propagation after diffraction from the twisted double slits. In addition, the delay levels of the $|\varphi_{06}\rangle$ and $|\varphi_{010}\rangle$ modes agree well with the theoretical prediction yet the error between theory and experiment for the $|\varphi_{012}\rangle$ mode is larger, as shown in Fig. 3(a) and (b) (but the delay ratio between the 1.2-m and 2-m propagation distances still agrees well with theory). This phenomenon is ascribed to the fact that the efficiency of the SLM decreases with the azimuthal indices $\ell$ when depicting a 0-2π phase shift with fewer pixels; for more details, see *Supplementary Materials*.

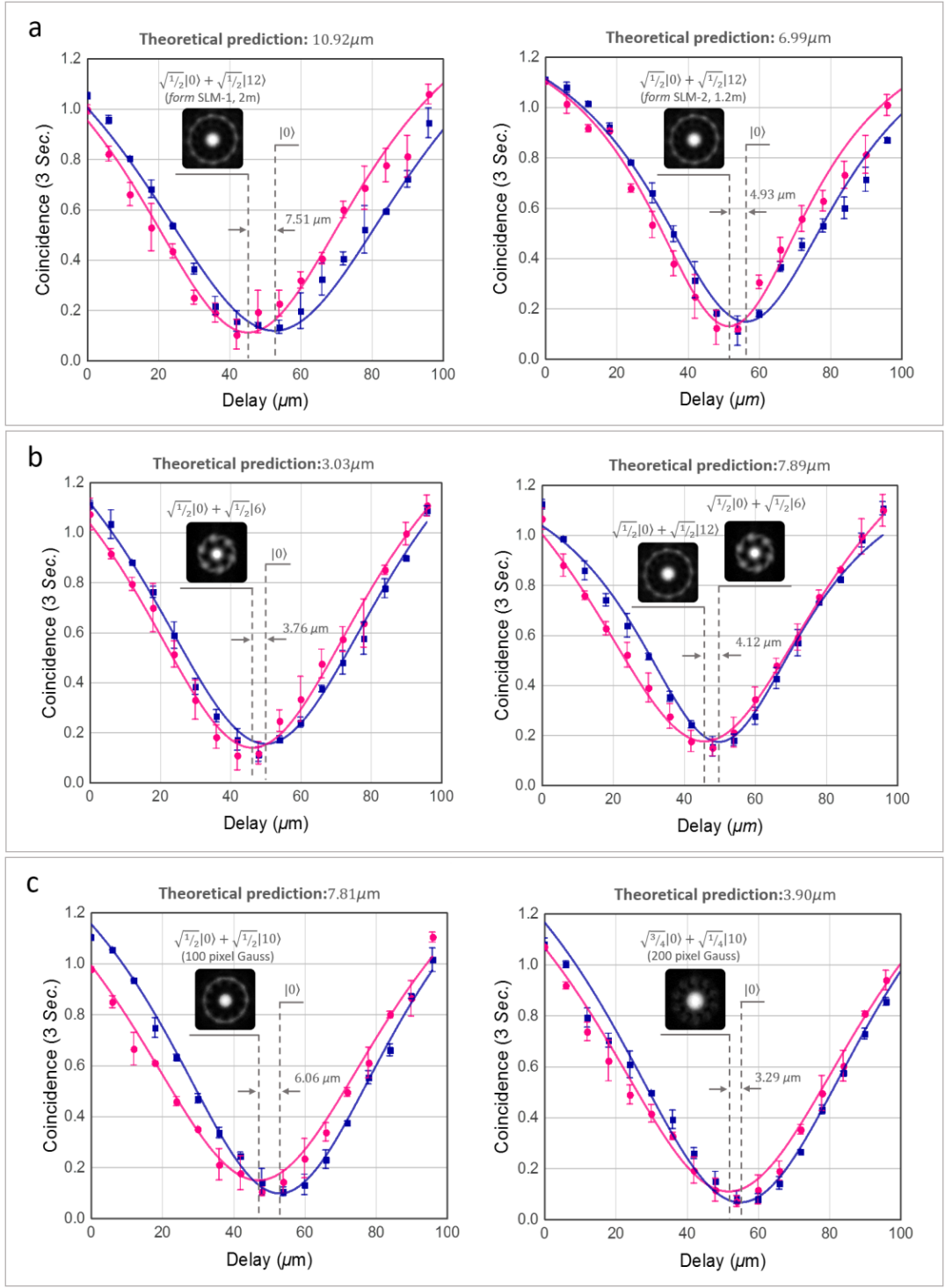

**Figure 3 | Experimental results of arriving photons' HOM interference.** The blue curves in all diagrams represent the arrival time for the Gaussian mode (SLMs operating in mirror mode). The grayscale images embedded in all curve diagrams represent transverse spatial wavefunctions of corresponding photons diffracting at 0.5 m from the SLM, and the diameters of the Gaussian slit in the holograms adopted in Fig. 3(a) and (b) are 100 pixels. The pixel size of the SLM is 6.4 $\mu$m, and the diameter of the incidence beam is approximately 1.5 mm. All photons are received from the 'Gaussian path'. **a,** Delays of $\sqrt{1/2}(|0\rangle+|12\rangle)$ photons arriving from SLM-1 and SLM-2. **b,** Difference between photons' arrival time for $|0\rangle$ and $\sqrt{1/2}(|0\rangle+|6\rangle)$, and for $\sqrt{1/2}(|0\rangle+|6\rangle)$ and $\sqrt{1/2}(|0\rangle+|12\rangle)$ modes. **c,** Delays of the superposition photons with different MWCs; the left and right diagrams correspond to $\sqrt{1/2}(|0\rangle+|12\rangle)$ and $\sqrt{3/4}|0\rangle+\sqrt{1/4}|10\rangle$ modes, respectively. The theoretical delays shown in a, b, and c are calculated from Eq. 5 using the data shown in Fig. 1.

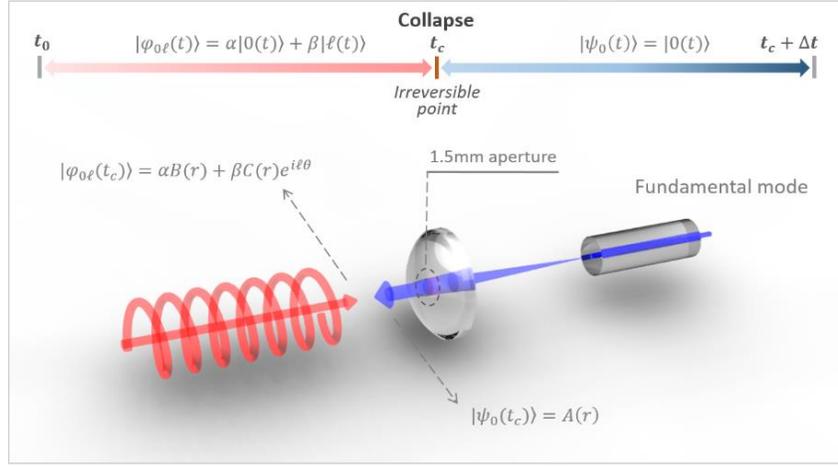

**Figure 4 | Collapse in fiber coupling.** The red light represents the incoming wavefunction $|\varphi_{0\ell}(t)\rangle$ that evolves into $|\varphi_{0\ell}(t_c)\rangle$ as it arrives at the surface of the coupling lens. The blue light represents the field-of-view (FOV) wavefunction emitted from the fiber collimator, which is a fundamental mode in the SMF and evolves into a Gaussian profile $|\psi_0(t_c)\rangle$ as it arrives at the surface of the coupling lens. At time $t_c$, according to Eq. 7, the partial incoming wave will be irreversibly coupled into the fiber and converted into the fundamental mode, i.e., a collapse process breaks the time-reversal symmetry before and after $t_c$.

The deterministic history of an unregistered photon revealed in this experiment challenges neither common sense nor causality. On the contrary, this result clarifies a long-held misunderstanding of the role of wavefunctions in the evolution of quantum entities and explains why causality can be held in the realistic interpretation. On one hand, a key point to comprehending this outcome is that we observed the arriving photons' particle nature $|0\rangle$, for which the obtained intensity signals (photon number) manifested no interference feature ($\alpha^2$). However, the arrival time recorded by fourth-order interference between signal and reference photons, or HOM interference, provides additional information for inquiry into the propagation history. On the other hand, this result is amazing because the observed photon seems to have definitely travelled along the Gaussian path $|0\rangle$, yet it indeed comes from both paths $|\varphi_{0\ell}\rangle$. First, the realistic interpretation of a wavefunction implies that the phenomena registered in collapse measurements do not represent the past of quantum entities. Due to the time-reversal invariance of wave dynamics, the observation in a measurement depends on the projection measurement between the incoming wavefunction $|\varphi_{0\ell}(t)\rangle$ and the field-of-view (FOV) wavefunction $|\psi_0(t)\rangle$ of the observation device (see Eq. 7 in Methods). Second, collapses violate the time-reversal symmetry of the wave dynamical process, and one cannot infer the previous state of a quantum entity from the result obtained in a collapse measurement via time-reversal calculations.

Specifically, as shown in Fig. 4, before the wavefunction $|\varphi_{0\ell}(t)\rangle$ collapses into $|0(t)\rangle$ or $|\ell(t)\rangle$ via a Gaussian-mode projection at time $t_c$, the evolution of individual photons obeys $|\varphi_{0\ell}(t)\rangle$, which forms the propagation history ($t_0 \sim t_c$) in free space. It is important to note that the delay obtained in the experiment cannot directly inform us of the photons' past ($t_0 \sim t_c$); to infer its history prior to the collapse ($t_c$), the knowledge of the experimental setup (such as propagation distance and phase mask of the SLM) is necessary. In other words, observations do not change or construct the history of quantum entities but determine their following fate; collapses provide an underlying principle for obtaining irreversible process from time-symmetric dynamics (Loschmidt's or the reversibility paradox)[29].

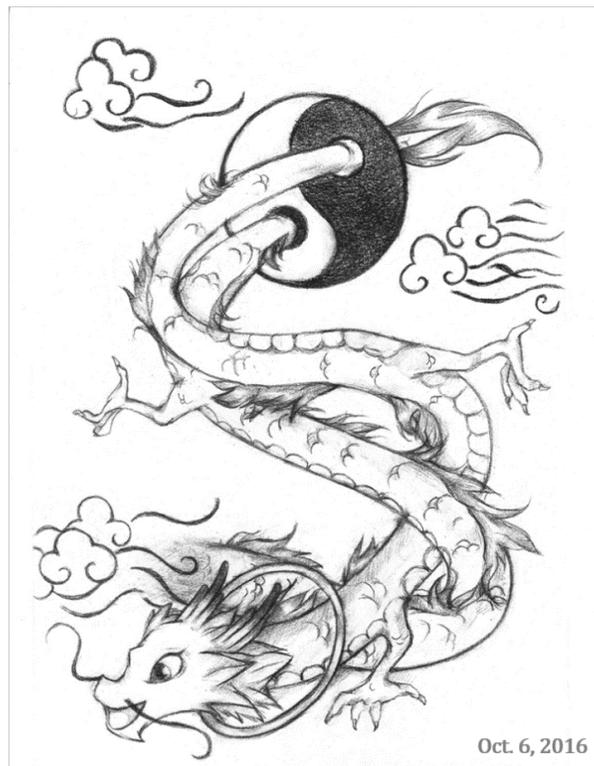

**Figure 5 | Quantum twisted Loong.** The tail and head of the Loong represent the particle nature of a photon; in contrast, the two coexisting bodies represent the wave nature. Here, the "Yin-Yang" corresponds to the twisted double slit, which transforms a photon's particle nature into wave nature. The ring (Qian-Kun Ring) drawn at the bottom represents an observation performed in one of the paths (projection measurement): Once a Loong is observed to come through the ring, the other body will instantly disappear. Now, imagine that observers standing at the bottom observes the outside through the ring; they will see that the Loong comes from the "Yin" (black) slit because the ring breaks the time-reversal symmetry of the dynamics.

In summary, we have presented a quantum twisted double-slit experiment, in which the nature of individual photons during their time in flight is revealed for the first time. The physical reality of their quantum states is verified, i.e., the wavefunction describes the true existence of individual photons after passing through the double slits, which is deterministic in evolution and predictable in measurements. In addition, the collapse of the wavefunctions breaks the time-reversal symmetry of wave dynamics, which determines the time progression of dynamics and maintains the causality in collapse measurements. These results highlight that quantum entities' existence and evolution are pre-existing and are not decided by God throwing dice; remind us that quantum mechanics has provided an underlying mechanism of obtaining irreversibility from a reversible system. Finally, we used a cartoon of the "quantum twisted Loong", shown in Fig. 5, to illustrate the propagation behavior of photons in a double-slit apparatus. Unlike the cartoon of "great smoky dragon" presented by Miller and Wheeler in 1983, where the body of the dragon is unknown and smoky[30], the body of the Loong is deterministic—two bodies coexist simultaneously before the collapse.

**References**

1. Bohr, N. The quantum postulate and the recent development of atomic theory. *Nature* **121,** 580 (1928).
2. Bohr, N. in *Quantum Theory and Measurement* (eds Wheeler, J. A. &Zurek, W. H.) 9–49 (Princeton University Press, 1984).
3. Einstein, A., Podolsky, B., and Rosen, N. Can quantum-mechanical description of physical reality be considered complete? *Phys. Rev.* **47,** 777 (1935).
4. Lindley, D. Focus: What's wrong with quantum mechanics? *Physics*, **16,** 10 (2005).
5. Ma, X. S., Kofler, J., & Zeilinger, A. Delayed-choice gedanken experiments and their realizations. *Rev. Mod. Phys.* **88** (1), 015005 (2016).
6. Lundeen, J. S., Sutherland, B., Patel, A., Stewart, C., & Bamber, C. Direct measurement of the quantum wavefunction. *Nature* **474** (7350)**,** 188-191 (2011).
7. Lundeen, J. S., & Bamber, C. Procedure for direct measurement of general quantum states using weak measurement. *Phys. Rev. Lett.* **108** (7)**,** 070402 (2012).
8. Aharonov, Y., and Zubairy, M. S. Time and the quantum: erasing the past and impacting the future. *Science* **307,** 875 (2005).
9. Wheeler, J. A. in *Mathematical Foundations of Quantum Theory* (eds Marlow, A.R.) 9–48 (Academic Press, 1978).
10. Hellmut, T., Walther, H., Zajonc, A. G. & Schleich, W. Delayed-choice experiments in quantum interference. *Phys. Rev. A* **35,** 2532–2541 (1987).
11. Baldzuhn, J., Mohler, E. & Martienssen, W. A wave–particle delayed-choice experiment with a single-photon state. *Z. Phys. B* **77,** 347–352 (1989).
12. Lawson-Daku, B. J. et al. Delayed choices in atom Stern–Gerlach interferometry. *Phys. Rev. A* **54,** 5042–5047 (1996).
13. Kim, Y. H., Yu, R., Kulik, S. P., Shih, Y. & Scully, M. O. Delayed 'choice' quantum eraser. *Phys. Rev. Lett.* **84,** 1–5 (2000).
14. Jacques, V. et al. Experimental realization of Wheeler's delayed-choice gedanken experiment. *Science* **315,** 966–968 (2007).
15. Jacques, V. et al. Delayed-choice test of quantum complementarity with interfering single photons. *Phys. Rev. Lett.* **100,** 220402 (2008).
16. Ionicioiu, R. & Terno, D. R. Proposal for a quantum delayed-choice experiment. *Phys. Rev. Lett.* **107,** 230406 (2011).

17. Ma, X. S. et al. Experimental delayed-choice entanglement swapping. *Nat. Phys.* **8,** 479–484 (2012).
18. Ma, X. S., et al. Quantum erasure with causally disconnected choice. *Proc. Natl. Acad. Sci. U.S.A.* **110,** 1221 (2013).
19. Tang, J. S., Li, Y. L., Li, C. F., & Guo, G. C. Realization of quantum Wheeler's delayed-choice experiment. *Nat. Photon.* **6,** 600 (2012).
20. Tang, J. S., Li, Y. L., Li, C. F., & Guo, G. C. Revisiting Bohr's principle of complementarity using a quantum device. *Phys. Rev. A* **88,** 014103 (2013).
21. Danan, A., Farfurnik, D., Bar-Ad, S., & Vaidman L. Asking photons where they have been. *Phys. Rev. Lett.* **111,** 240402 (2013).
22. Bouchard, F., Harris, J., Mand, H., Boyd, R. W., & Karimi, E. Observation of subluminal twisted light in vacuum. *Optica*, *3*(**4**)**,** 351-354 (2016).
23. Giovannini D., *et al*. Spatially structured photons that travel in free space slower than the speed of light. *Science*, **347** (6224)**,** 857-860 (2015).
24. Allen, L. *et al*. Orbital angular momentum of light and the transformation of Laguerre-Gaussian laser modes. *Phys. Rev. A* **45** (11)**,** 8185 (1992).
25. Torres, J. P. and Torner, L. *Twisted photons: applications of light with orbital angular momentum* (John Wiley & Sons, 2011).
26. Yao, A. M., & Padgett, M. Orbital angular momentum: origins, behavior and applications. *Adv. Opt. Photon.* **3,** 161-204 (2011).
27. Zhou, Z. Y. *et al*. Generation and reverse transformation of twisted light by spatial light modulator. *arXiv*:1612.04482 [physics.optics] (2016).
28. Padgett, M. J. & Allen, L., The poynting vector in Laguerre–Gaussian laser modes. *Opt. Commun.* **121,** 36–40 (1995).
29. J. Loschmidt, Sitzungsber. Über den Zustand des Wärmegleichgewichts eines Systems von Körpern mit Rücksicht auf die Schwerkraft. *Kais. Akad. Wiss. Wien, Math. Naturwiss.* Classe **73**, 128–142 (1876).
30. Miller, W. A., & Wheeler, J. A. In *Proceedings of the international symposium foundations of quantum mechanics in the light of new technology*. (1983).

## Acknowledgments

We thank Prof. Jeff S. Lundeen from University of Ottawa for fruitful discussions. A special thank and bless to Yi-Fan Gao, a smart and pretty girl from Harbin Institute of Technology, for drawing the cartoon of quantum twisted Loong. This work is supported by the National Natural Science Funds for Distinguished Young Scholar of China (Grant No. 61525504); the National Natural Science Foundation of China (Grant Nos. 11574065, 11604322, 61275115, 61378003, 61435011, 61605194); China Postdoctoral Science Foundation (Grant No. 2016M590570); the Fundamental Research Funds for the Central Universities No. 11604322), and the Key Programs of the Natural Science Foundation of Heilongjiang Province of China (Grant No. ZD201415).

## Author contributions

Z.-H. Z. conceived and designed the experiment, Z.-Y. Z developed the theory. The experimental work was conducted by Z.-Y. Z and Z.-H. Z. with assistance from S.-L. L., Y.-H. L, and S. S. Moreover, D.-S. D, L.-X. C., and W. G. analyzed the results. Z.-H. Z. wrote the manuscript and all authors contributed to discussions during its preparation. G.-C.G and B.-S.S. supervised the project.

## Competing financial interests

The authors declare no competing financial interests.

## Methods

### Theoretical frame

**Slowing down twisted light.** The transverse wavefunction of twisted light after generation can be expressed as

$$E(r,\theta,z) = \frac{i^{\ell+1}}{2\lambda z}\sqrt{\frac{2}{\pi}}\frac{1}{w_0}\exp(-ikz)\exp(-i\ell\theta)\exp(-\frac{ik}{2z}r^2)\frac{b^{\ell}}{\varepsilon^{1+\ell/2}}F(\frac{\ell}{2},\ell+1,\frac{b^2}{\varepsilon}), \quad (1)$$

where $F(\alpha,\beta,z)$ is a hypergeometric function and $b = kr/2z, \varepsilon = 1/w_0^2 + ik/2z$ (for more details, see *Supplementary Materials*). The relationship between the wave vector $k_0$ and the related components in cylindrical coordinates $\{r,\varphi,z\}$ can be expressed as $k_0^2 = k_r^2 + k_\varphi^2 + k_z^2$, and the axial-group velocity $v_{g,z}$ can be expressed as[23]

$$v_{g,z} = \frac{ck_z}{k_0} = c\left(1 + \frac{\langle \Delta k_\perp^2 \rangle}{2k_0^2}\right)^{-1}, \quad (2)$$

where $\langle \Delta k_\perp^2 \rangle = -\langle \varphi | \nabla_\perp^2 | \varphi \rangle$, corresponding to the transverse component of the total wave vector, and $\nabla_\perp^2 = \frac{1}{r}\frac{\partial}{\partial r}(r\frac{\partial}{\partial r}) + \frac{1}{r^2}\frac{\partial^2}{\partial\varphi^2}$ is the Laplacian operator in the transverse plane, i.e., $k_r$ and $k_\varphi$. The accumulated delay in the optical path that results from the reduction in group velocity can be expressed as[23]

$$\tau(z) = \int_0^z (1 - \frac{c}{v_{g,z}(z)})dz. \quad (3)$$

The phase of an LG mode (twisted light) superposition can be expressed as

$$|\varphi\rangle = \sum_{\ell=0} c_\ell |\varphi\rangle_\ell, \sum_{l=0}|c_\ell|^2 = 1. \quad (4)$$

Thus, the wave vector in the transverse plane is $\langle \hat{k}_\perp^2 \rangle = \sum_{m=0}\sum_{n=0} c_m^* c_n \langle \varphi|_m \nabla_\perp^2 |\varphi\rangle_n = -\sum_{n=0}|c_n|^2 \langle \varphi|_n \nabla_\perp^2 |\varphi\rangle_n$, and according to Eq. 2, we obtain the axial-group velocity as

$$v_{g,z} = c(1 + \frac{1}{2k_0^2}\sum_{n=0}|c_n|^2 \langle \varphi|_n \nabla_\perp^2 |\varphi\rangle_n). \quad (5)$$

This result indicates that the delay of a superposition mode should be an 'intermediate delay' manifesting a linear superposition of different modes. According to Eq. 4, the accumulated delay of $|\varphi_{0\ell}\rangle = \alpha|0\rangle + \beta|\ell\rangle$ ($|\alpha|^2 + |\beta|^2 = 1$) after propagating a distance of z is

$$|\beta|^2 \tau_\ell(z), \quad (6)$$

where $\tau_\ell(z)$ is the accumulated delay of the $|\ell\rangle$ mode as a function of propagation distance.

**Collapse in fiber coupling.** The mode selection of the SMF for the superposition state can be described as follows.

First, the SMF can support only the fundamental mode (Gaussian mode), as shown in Fig. 4, which can be converted into a spatial Gaussian mode by a fiber collimator, representing the field-of-view (FOV) wavefunction $|\psi_0(t)\rangle$ of the fiber coupling. The FOV wavefunction evolves into $|\psi_0(t_c)\rangle = A(r)$ (a Gaussian profile) at the surface of the coupling lens, where $\iint |A(r)|^2 rdrd\theta = 1$. Based on the time-reversal invariance of wave equations, the largest FOV that can be observed via the collimator is the time-reversed wave of the FOV wavefunction, in the form of $A^*(r)$ on the surface of the coupling lens.

Second, the incoming wavefunction $|\varphi_{0\ell}(t)\rangle$ can be expressed as $|\varphi_{0\ell}(t_c)\rangle = \alpha B(r) + \beta C(r)e^{i\ell\theta}$ when arriving at the surface of the coupling lens, where $\iint |B(r)|^2 rdrd\theta = 1$ and $\iint |C(r)e^{i\ell\theta}|^2 rdrd\theta = 1$. Now, consider a scenario in which the SMF

is used to receive the incoming wave: the field received into the SMF can be expressed as the projection measurement between $|\varphi_{0\ell}(t_c)\rangle$ and $|\psi_0(t_c)\rangle$:

$$\Phi(r,\theta) = \iint (\alpha B(r) + \beta C(r)e^{i\ell\theta})A^*(r)rdrd\theta \Rightarrow \iint \alpha B(r)A^*(r)rdrd\theta . \tag{7}$$

Note that Eq. 7 *represents an irreversible collapse process* because the field received into the fiber is the fundamental mode and its time-reversed wave is the Gaussian mode, not the superposition state $|\varphi_{0\ell}\rangle$. The coupling efficiency can be expressed as

$$\eta = \left|\iint \Phi(r,\theta)rdrd\theta\right|^2 = \alpha^2 \left|\iint B(r)A^*(r)rdrd\theta\right|^2 , \tag{8}$$

where $\left|\iint B(r)A^*(r)rdrd\theta\right|^2$ depends on the degree of transverse similarity between $A^*(r)$ and $B(r)$ and $\eta$ achieves $\alpha^2$ only when $B(r) = A^*(r)$ and in the absence of specular reflection.

**Paradox in photons' arrival time.** The distinguishability of Gaussian-mode filtering $D$, performed by the SMF, is defined as

$$D = \frac{|N_G - N_{LG}|}{N_G + N_{LG}} , \tag{9}$$

where $N_G$ and $N_{LG}$ are signal photons received from a Gaussian mode and an LG mode, respectively, and in the experiment, $D \approx 1$ (for details, see *Supplementary Materials*).

In fiber coupling, the incoming wavefunction $|\varphi_{0\ell}(t)\rangle = \alpha|0(t)\rangle + \beta|\ell(t)\rangle$ will collapse into $\sqrt{D}|0\rangle + \sqrt{1-D}|\ell\rangle \approx |0\rangle$. Now, consider the following two either-or cases:

1. In a projection (collapse) measurement, if the photon behaves as neither a particle nor a wave before the measurement and the propagation history is formed by the observed state, or in the Copenhagen viewpoint, there shall be no delay because the propagation history ($t_0 \sim t_c$) is formed by the $|0(t)\rangle$ state;

2. In contrast, if the nature of the photon is pre-existing in projection (collapse) measurements as a form of its wavefunction, then the 'intermediate delay' obeying Eq. 6 will still emerge because the propagation history ($t_0 \sim t_c$) in free space was formed by $|\varphi_{0\ell}(t)\rangle$ before its collapse into the $|0(t)\rangle$ state at time $t_c$.

## Experimental setup

In the experiment, to measure the arrival time of single photons with fs-level precision, a 1mm type-II periodically poled potassium titanyl phosphate (PPKTP) crystal is used to generate entangled photon pairs with a 160fs duration via spontaneous parametric down-conversion (SPDC). One of the entangled photons is used as a signal and directed into the double-slit apparatus consisted of two SLMs and a fiber coupler, the other acts as a herald and transmits through the reference way containing fibers and an adjustable air-gap, and then a Hong-Ou-Mandel (HOM) interference between the entangled photon pair is performed to obtain the variation in arrival time of the signal photon. For more details, see *Supplementary Materials*.

# Supplementary Materials for

# Revealing photons' past via quantum twisted double-slit experiments

Zhi-Yuan Zhou[1,2], Zhi-Han Zhu[3,*], Shi-Long Liu[1,2], Yin-Hai Li[4], Shuai Shi[1,2], Dong-Sheng Ding[1,2], Li-Xiang Chen[5], Wei Gao[3], Guang-Can Guo[1,2], and Bao-Sen Shi[1,2,*]

[1]CAS Key Laboratory of Quantum Information, USTC, Hefei, Anhui 230026, China.
[2]Synergetic Innovation Center of Quantum Information & Quantum Physics, University of Science and Technology of China, Hefei, Anhui 230026, China.
[3]Da-Heng Collaborative Innovation Center for Science of Quantum Manipulation & Control, Harbin University of Science and Technology, Harbin 150080, China.
[4]Department of Optics and Optical Engineering, University of Science and Technology of China, Hefei, Anhui 230026, China.
[5]Department of Physics and Collaborative Innovation Center for Optoelectronic Semiconductors and Efficient Devices, Xiamen University, Xiamen 361005, China.
*e-mail: zhuzhihan@hrbust.edu.cn; drshi@ustc.edu.cn;

## I. Theoretical calculation

Here, for accurately calculating $v_{g,z}$ reduction, we consider the whole evolution of twisted light since be generated from a SLM. The transverse wavefunction of a perfect Gaussian beam with a beam waist $w_0$ can be expressed as $\sqrt{2/\pi}\exp(-r^2/w_0^2)$. When it loads a helical phase $e^{-i\ell\theta}$ at the surface of SLM or the transformation plane $\{z_0, t_0\}$, it will experience a phase transition and become

$$E_0(r_0,\theta_0) = \sqrt{2/\pi}\exp(-r_0^2/w_0^2)\exp(-i\ell\theta_0). \tag{S1}$$

The following evolution can be derived from the Collins Diffraction Integral Equation with the initial condition of $\{E_0, z_0, t_0\}$, as shown below

$$E_1(r_1,\theta_1,z_1) = \frac{i}{\lambda B}\exp(-ikz_1)\int_0^{2\pi}\int_0^{\infty} E_1(r_0,\theta_0)\exp\left\{-\frac{ik}{2B}\times[Ar_0^2 - 2r_1 r_0\cos(\theta_1-\theta_0) + Dr^2]\right\}r_0 dr_0 d\theta_0, \tag{S2}$$

where $z_1$ is the propagation distance since generation, and the ABCD matrix for free space propagation is

$$\begin{pmatrix} A & B \\ C & D \end{pmatrix} = \begin{pmatrix} 1 & z \\ 0 & 1 \end{pmatrix}. \tag{S3}$$

After integrating, we obtain an analytical solution for describing the wavefunction of twisted light during the diffraction process, as shown below (more details see *Ref.* 29)

$$E_1(r_1,\theta_1,z_1) = \frac{i^{\ell+1}}{2\lambda z_1}\sqrt{\frac{2}{\pi}}\frac{1}{w_0}\exp(-ikz_1)\exp(-i\ell\theta_1)\exp(-\frac{ik}{2z_1}r_1^2)\frac{b_1^{\ell}}{\varepsilon_1^{1+\ell/2}}F(\frac{\ell}{2},\ell+1,\frac{b_1^2}{\varepsilon_1}), \tag{S4}$$

where $F(\alpha,\beta,z)$ is the Hypergeometric function, and $b_1 = kr_1/2z_1$, $\varepsilon_1 = 1/w_0^2 + ik/2z_1$. Figure S1 shows the transverse profile of the twisted light upon propagation.

The differential of the field in the transverse directions

1. $-4g(\ell+1)r^\ell \exp(-gr^2)F(\ell/2,\ell+1,fr^2)$;

2. $-4gf\dfrac{l}{\ell+1}r^{\ell+2}\exp(-gr^2)F(\ell/2+1,\ell+2,fr^2)$;

3. $4f\dfrac{l}{\ell+1}r\ell\exp(-gr^2)F(\ell/2+1,\ell+2,fr^2)$;

4. $f^2\dfrac{\ell}{\ell+1}r^{\ell+2}\exp(-gr^2)F(\ell/2+2,\ell+3,fr^2)$;

5. $4g^2 r^{\ell+2}\exp(-gr^2)F(\ell/2,\ell+1,fr^2)$,

where $g = \dfrac{k}{2}\dfrac{z_R+iz}{z^2+z_R^2}, f = \dfrac{k}{2}\dfrac{z_R(z_R+iz)}{z(z^2+z_R^2)}$, and according to Eq. 2 shown in Method, the calculated slowing downing of the twisted light are shown in Fig. 1 (in the text) and the right inset of Fig. S1. Note, these results are calculated from ideal conditions, i.e., based on the twist light birth from a perfect Gaussian light a perfect helical phase as shown in Eq. S1.

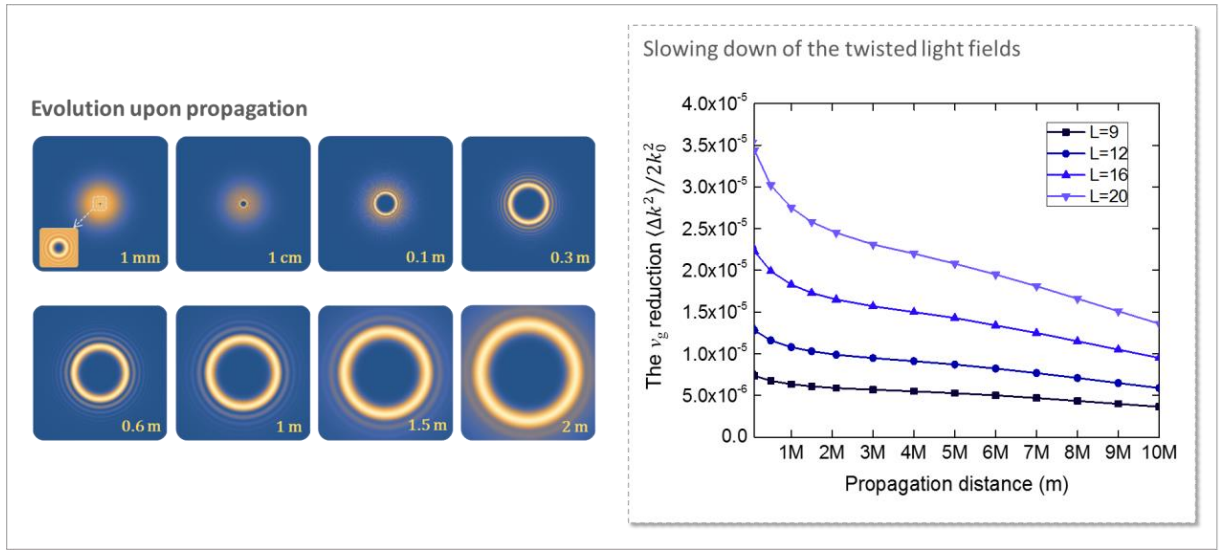

**Figure S1 | Theoretical calculation results.** The left top inset is the transverse spatial wavefunction of the realistic twisted light upon propagation (here $\ell=10$, $w_0=1.5$ mm), and the left bottom inset shows the transverse spatial wavefunction at 1 mm propagation distance with different zoom scales ($\ell=10$). The right inset is the calculated slowing down of the twisted light versus propagation distance. ($w_0=1.5$ mm).

## II. Details of beam profile and mode filtering

In the experiment, the twisted light was converted from Gaussian light by phase-only type SLM. As theoretical analyzed in above, the vortex inlayed at center of the donut will gradually enlarge upon propagation. Figure S2 shows the beam profiles of corresponding twisted light ($\ell$ =6, 8, 10 and 12) after propagating 1m. It can be see that the inner diameter d of them are all much larger than the aperture of the fiber collimator (1.5 mm). That is to say, the mode orthogonality or distinguishability should be 100% in this experimental setup (directly receiving and no reversed transformation reconverting them back to Gaussian light). It can be note, there is still few light can be coupled into the SMF with a coupling efficiency of ~1%, however, this few light is not real twisted light, but unconverted Gaussian light that directly reflect from the SLM. This is because that the efficiency of SLM is not 100%, and it decreases with $\ell$ due to depicting 0-2 π phase shift with fewer pixels. As shown in Fig. S2, the brightness of light spot inlayed at the donut's center increase with $\ell$, which consist with the increase of the fiber coupling efficiency (FCE). Nevertheless, these few unconverted light will combine into the Gaussian part of a superposition state $|\varphi_{0\ell}\rangle$, and the mode distinguishability in the experiment are still almost 100%, i.e., $D \approx 1$.

Additionally, we should note that the helical phase $e^{-i\ell\theta}$ loaded in light by a SLM is not a perfect continuous 0-2 π phase shift, but a step-type phase shift depicted by pixels of SLM. Obviously, the smoothness degree of this step-type phase shift will drastically reduce with increase of $\ell$ (depicting 0-2 π phase shift with fewer pixels). Comparing beam profile shown in Fig. 2S (a) and (d), it can be see that the mode of $\ell = 6$ is much better than mode of $\ell = 12$. Thus, the slowing down obtained in experiment shall more consist with the theoretical prediction in case of using a smaller $\ell$.

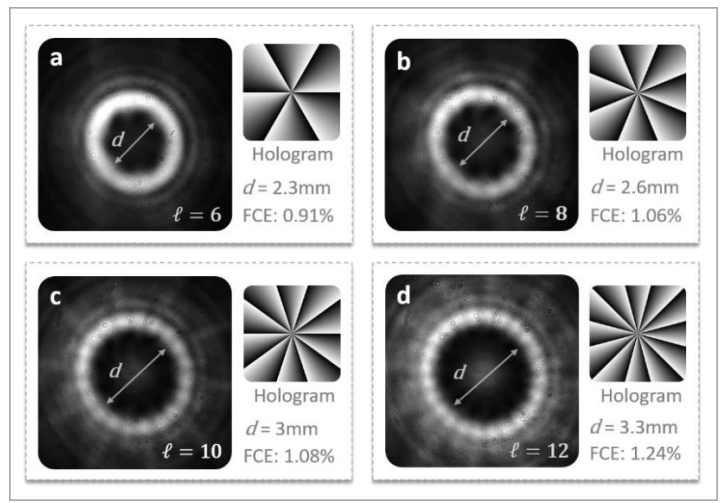

**Figure S2 | Beam profile of twisted light after propagating 1m since be generated via SLM.** The parameter *d* and fiber coupling efficiency (FCE) represent the inner diameter and fiber coupling efficiency of corresponding twisted light ($\ell$ =6, 8, 10 and 12) after propagating 1m, respectively.

## III. Experimental details

To obtain a high stability and brightness TEM$_{00}$-mode SPDC pump source, a frequency doubling (FD) in cavity technique is adopted in the experiment, as shown in Fig. S1. The type-I PPKTP crystal (Raicol Crystals) with dimensions of 1×2×10 (mm); the crystal was periodically poled with periodicity of 3.15 μm to obtain quasi phase matching (QPM) for second harmonic generation (SHG) from 795 to 397.5 nm. Both crystal end faces have anti-reflective coatings at the wavelengths of 397.5 and 795 nm. The crystal is x-cut for pump beam propagation along the x-axis of the crystal. The crystal temperature is controlled by a semiconductor Peltier device with stability of ±2 mK.

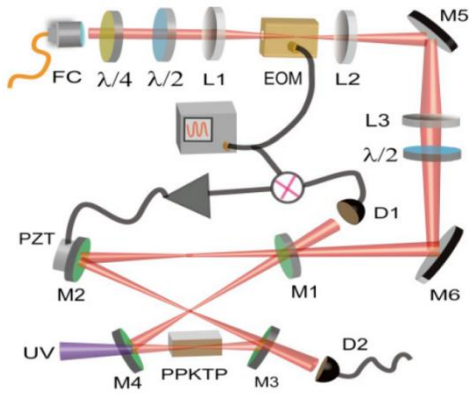

**Figure S3 | Schematic of high stability and brightness SPDC pump source generation.**

Figure S3 shows the detail setup. The FD pump laser is a CW external-cavity laser with a Tapered Amplifier tuned to a wavelength of 795 nm. Part of the laser beam is coupled to a single-mode fiber to produce a better spatial mode to pump the SHG cavity. To ensure the stability of the pump laser, a Faraday isolator is placed before the fiber coupler (FC) to prevent backscattering. The pump polarization is controlled by two wave plates that are placed after the fiber output port. The beam then enters an electro-optical modulator (EOM) to create sidebands to lock the cavity to the TEM$_{00}$ mode using the Pound–Drever–Hall method. After modulation by the EOM, the pump light's mode is matched to the cavity by a focusing system consisted of two lenses, L2 and L3. The reflected light from the input mirror M1 is detected by a fast detector D1, and the signal from D1 is mixed with the RF modulation signal from the EOM by a common signal generator. The mixed signal passes through a low-pass filter, and an error signal is generated. The error signal is then processed using a home-made servo control and a high-voltage amplifier to control the piezoelectric transducer that is attached to mirror M2. The detector D2 is used to monitor the status of the cavity. The generated SH beam is filtered using a dichromatic mirror. The ring cavity is formed by four mirrors, where the input coupling mirror M1 has a transmittance of 5% at 795 nm, and mirrors M2–M4 have highly reflective coatings at 795 nm (R>99.9%). The output mirror M4 alone has a transmittance of 80% at 397.5 nm. Mirrors M3 and M4 are concave mirrors with 80 mm curvatures. The geometry of the cavity was designed using the ABCD matrix method. At last, the high-stable UV laser is coupled into a single-mode fiber to produce a perfect Gaussian mode. More details are shown in *COL* **12**(11), 111901 (2014).

In the experiment, to ensure a better optical path stability, the cavity is isolated from the table vibration and all fibers are covered by tinfoil paper to insulate heat exchange from the environment. Figure S4 shows the measured HOM interferences before and after the heat and vibration isolation. Moreover, photon pairs generated from a 2.5 mm type-II PPKTP with a 400 fs duration are also adopted in early stage of the experiment, corresponding results are shown in Fig. S5 which presents the same regularity shown in the main body of the paper.

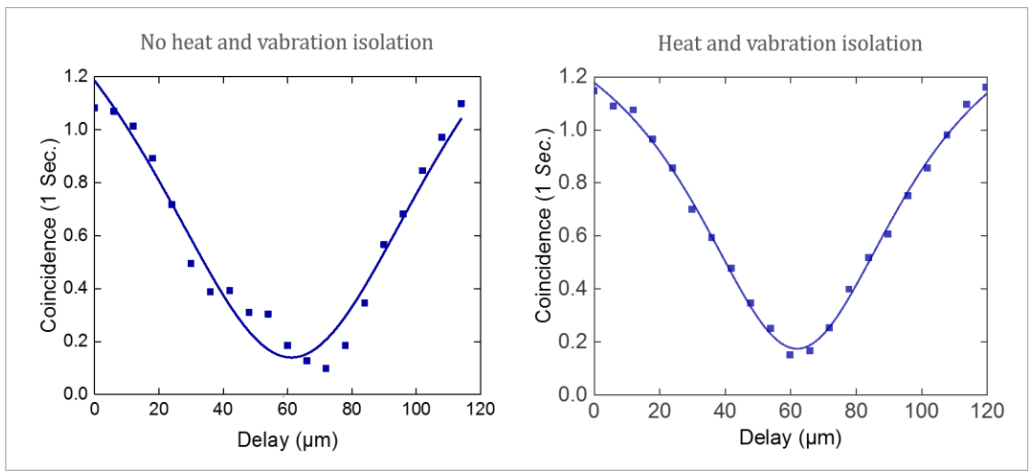

**Figure S4 | HOM interferences (Gaussian photon pairs) before (left) and after (right) the heat and vibration isolation.**

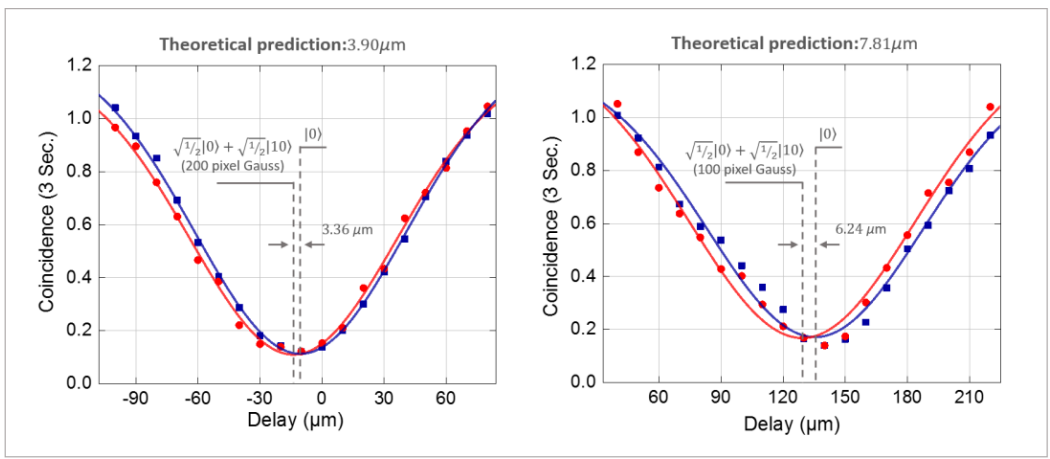

**Figure S5 | HOM interferences with 400fs duration photon pairs.** Delay of the superposition photons with different path weight ratios, the left and right diagrams correspond to photons transformed from holograms with 200 pixels and 100 pixels diameters of Gaussian part, respectively